# Pricing Mechanism in Information Goods

Xinming Li[1], Huaqing Wang[2]

1. School of Management, Xi'an Jiaotong University (China)
2. School of Business, Emporia State University (USA)

**Abstract**

We study three pricing mechanisms' performance and their effects on the participants in the data industry from the data supply chain perspective. A win-win pricing strategy for the players in the data supply chain is proposed. We obtain analytical solutions in each pricing mechanism, including the decentralized and centralized pricing, Nash Bargaining pricing, and revenue sharing mechanism.

## 1. Introduction and Literature Review

"Big Data" has recently become the focus of academic and corporate investigation (Wamba, 2015). In practice, the economic value of big data is reflected in the success of many Internet companies, from search engines to social media sites and data repositories which routinely sell this information (Gkatzelis, Aperjis & Huberman, 2015). In academic, a new scientific paradigm is born as data-intensive scientific discovery, which involves a large number of fields and sectors, ranging from economic and business activities to public administration, from national security to scientific researches in many areas (Chen & Zhang, 2014). In the big data era, data have been seen as a new asset and been sold in the market. For example, recently there emerges a new kind of startups specialized in selling data. These emerging startups claim themselves as "data bank" in big data industry, (e.g., Shujutang, http://www.datatang.com) and their value proposition is to provide raw data for the whole big data industry, allowing other participants make the best use of the data. Data pricing mechanism, which involves the profit allocation and incentive issues, is very fundamental in big data industry, yet the research in this area has not been given sufficient consideration.

To fill this research gap, we investigate the data pricing mechanism from data supply chain perspective. Along the big data value chain from upstream to downstream (Curry, 2016), there are basically two types of representative firm. One is the upstream data provider who provides raw data for the whole data industry. The other one is the downstream application provider who provides end-users with data-based application. The data provider, application provider, and the end-users consist the data supply chain. Therefore, this paper, from the perspective of data supply chain, investigates the pricing and coordination issues among participants in the data transaction.

We consider the new features of data product or service (thereafter we use data product throughout this paper). Although data product can be seen as a kind of information goods (Sarvary and Parker 1997; Raghunathan & Sarkar, 2016), they differ from the traditional information goods. First, data provider can collect end-users' data (referred to as feedback data from end-users), which have value for the data provider. For example, Internet companies offer free service to attract users and then monetize users' personal data (Li, Li, Miklau & Suciu, 2012). Nowadays, through API (Application Programming Interface), it is easy for the data provider to collet end-users' data. Second, the value of data is context-dependent, which means the same data product might have different value in different application context or for different end-users. Therefore, we consider the uncertainty of data value and feedback data from end-users in this study.

Our contribution to the existing related literature is two-fold. First, we add to the existing information goods pricing literature. The information goods pricing has been studied extensively. Sarvary & Parker (1997) study how information goods' marketing strategies should be different with traditional products. Sundararajan (2004) analyzes optimal pricing for information goods under incomplete information. Chen & Seshadri (2007) consider both information goods development and pricing issues. Different from this stream of pricing papers that are analyses from the seller-side, our research analyzes the pricing issue from the perspective of supply chain and addresses the interaction between upstream data provider and downstream application provider. More related to our paper is the recent research in private data pricing in the context of big data. Considering the individual's privacy concern, Li, Li, Miklau & Suciu (2012) propose a theoretical framework for assigning prices to queries in order to compensate the data owners for their loss of privacy. Gkatzelis, Aperjis & Huberman (2015) investigate the private data pricing under sellers' heterogeneous risk averse attitude and privacy concern. While these papers focus on the individual's private data pricing, our paper investigates the data used as a commercial product. Second, our closed-loop data supply chain differs from the traditional product closed-loop supply chain, which is mainly about reverse logistic (e.g., Dowlatshahi 2000; Wen et al. 2018). However, in our closed-loop data supply chain, the logistics issue does not exist and the main concern is business value of data. There is extensive literature on product closed-loop supply chain. We refer to Govindan Soleimani & Kannan (2015), which gives a comprehensive review.

## 2. Model Setup and Assumptions

We establish our data supply chain based on the reality of big data industry, and the data supply chain reflects the core value chain in the big data value chain in Curry (2016). The data supply chain consists an upstream data provider, a downstream application provider, and end-users, as Figure 1 shows.

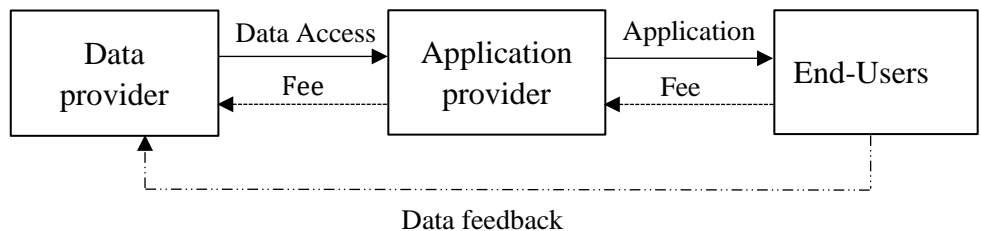

Figure 1. Closed-Loop Data Supply Chain Model

This data supply chain is a representative one in big data industry. Specifically, the data provider represents the upstream firm in the data value chain, which provides raw data for other firms that can make the best use of the data. The application provider represents the downstream firm in the data value chain, which serves the end-user in various industries, such as marketing, finance, retailer, and so on. Unlike the traditional supply chain, the data supply chain is close-looped and the data provider in the chain, through API, can collect data from the end-users. For example, a provider of navigation application uses data provider's database of digital map in its application and as a result, as the end-users use the navigation, the data provider gets the end-users' data in regard to their traveling behavior.

**The data provider**: the data provider offers the data (e.g., database access) of potential value $v$, and charge the unit usage price $w$. Since each end-user has some value for the data provider (according to Li, Li, Miklau & Suciu (2012), a recent study by JPMorgan Chase found that each unique user is worth approximately $4 to Facebook and $ 24 to Google), thus we assume $d$ as the value of each end-user for data provider. The data provider needs the downstream application provider to distribute the data, but cannot observe the realized value of application. In order to build a concise and tractable model, similar to Bhaskaran & Krishnan (2009), we model the value uncertainty as $\tilde{r}v$, where $\tilde{r}$ is uniformly distributed between $r$ and 1 ($\tilde{r} \sim U[r, 1]$). As $r$ increases, the data value uncertainty decreases.

**The application provider**: based on the data from the data provider, the application provider offers the end-users with application of value $\tilde{r}v$, and then the expected value of the application is $\frac{1+r}{2}v$. The application provider charges the end-users unite usage price $p$.

**The end-user**: we normalized the size of the end-users to unit one. End-users have heterogeneous willingness to pay for the application. Let $\theta$ denote the users' unit willingness to pay for the application value, and $\theta$ is uniformly distributed on $[0,1]$. Thus, the expected utility of the type $\theta$ consumer is $U = \theta \frac{1+r}{2} v - p$. The end-users use the application only if their expected utilities are non-negative. Solving for $\theta$ when $U = 0$, we can get the indifferent end-user $\theta^* = \frac{2p}{(1+r)v}$, and therefore the size of the end-user base (demand) is $D = \frac{(1+r)v - 2p}{(1+r)v}$.

We also make the following assumptions:

**Assumption 1.** Following the research in information goods, the information goods' variable cost is neglected and fixed cost is sunk. (see Sarvary & Parker, 1997; Sundararajan, 2004; Zhang, Nan, Li & Tan, 2016).

**Assumption 2.** Data provider and application provider have equal bargaining power, and they are risk-neutral.

**Assumption 3.** The distribution of the value is common knowledge for both data provider and application provider.

## 3. The Data Pricing Mechanisms

Following the standard analysis in supply chain literature, we begin with the decentralized and centralized pricing mechanism. Since the interactive pricing mechanisms, whereby the price is not solely determined by the seller but rather developed through the interaction between sellers and buyers, is widely used (Kannan & Kopalle 2001), we then examine the bargaining pricing mechanisms. Bargaining is commonly used in supply chain or channel context (Wu, Baron & Berman, 2009; Du, Nie, Chu & Yu, 2014; Iyer & Villas-Boas, 2003). Finally, we compare the performances of these pricing mechanisms.

## 3.1 Decentralized and Centralized Pricing Model

In the decentralized data supply chain, the data provider directly charges the application provider unite usage price w, and the application provider charges the end-users a price of p for the usage of its application. We model the interaction between the two players as a two-stage game, with the data provider selecting the wholesale price w in the first stage, and the application provider choosing the selling price p in the second stage. We characterize the equilibrium of the game by solving the application provider's problem first. For a given w, the application provider's optimization problem is given by:

$$\max_{p} \pi_2^D = (p - w)\frac{(1+r)v - 2p}{(1+r)v} \tag{1}$$

The profit function is concave in $p$ ($\frac{\partial^2 \pi_2^D}{\partial p^2} = -\frac{4}{(1+r)v} < 0$), and therefore the first-order condition is sufficient to characterize the optimal best response function $p^D = \frac{1}{4}(v + rv + 2w)$.

In the first stage, the data provider solves the following problem while taking into account the application provider's reaction:

$$\max_{w} \pi_1^D = (w + d)\frac{(1+r)v - 2p}{(1+r)v} \tag{2}$$

By solving for first-order condition, we get $w = \frac{1}{4}(v + rv - 2d)$, and the second-order condition is satisfied $\frac{\partial^2 \pi_1^D}{\partial w^2} = -\frac{2}{(1+r)v} < 0$, and thus $w = \frac{1}{4}(v + rv - 2d)$ is unique maximum point.

**Lemma 1.** Under the condition $v > \frac{2d}{3(1+r)}$, the equilibrium prices are $w = \frac{1}{4}(v + rv - 2d)$ and $p^D = \frac{1}{8}(3(1+r)v - 2d)$; the expected profits of data provider, application provider, and data supply chain are $\pi_1^D = \frac{(2d+v+rv)^2}{16(1+r)v}$, $\pi_2^D = \frac{(2d+v+rv)^2}{32(1+r)v}$, $\Pi^D = \pi_1^D + \pi_2^D = \frac{3(2d+v+rv)^2}{32(1+r)v}$, and the consumer surplus is $CS^D = \frac{(2d+v+rv)^2}{64(1+r)v}$.

**Proof.** Substituting the data provider's price decisions $w = \frac{1}{4}(v + rv - 2d)$ to application provider's price decision, we get $p^D = \frac{1}{8}(3(1+r)v - 2d)$. Thus, by substituting these price decisions back into the profit functions, we can get the expected profits for data provider and application provider. The consumer surplus is $CS^D = \int_{\theta^*}^{1} (\theta \frac{1+r}{2} v - p) \, d\theta = \frac{(2d+v+rv)^2}{64(1+r)v}$. The demand constraint $0 < \frac{(1+r)v - 2p}{(1+r)v} \leq 1$ requires $v > \frac{2d}{3(1+r)}$. ∎

According to Lemma 1, we can get the following proposition.

***Proposition 1.*** *When the value of feedback data from end-users is relatively significant ($\frac{1+r}{2} v < d < \frac{3(1+r)}{2} v$), all profits ($\pi_1^D, \pi_2^D, \Pi^D$) are negatively related to $r$. No matter what data value uncertainty ($r$), all profits ($\pi_1^D, \pi_2^D, \Pi^D$) increase with the value of end-users' feedback data ($d$).*

**Proof.** $\frac{\partial \pi_1^D}{\partial r} = \frac{(1+r)^2 v^2 - 4d^2}{16(1+r)^2 v}$ ; $\frac{\partial \pi_2^D}{\partial r} = \frac{(1+r)^2 v^2 - 4d^2}{32(1+r)^2 v}$; $\frac{\partial \Pi^D}{\partial r} = \frac{3((1+r)^2 v^2 - 4d^2)}{32(1+r)^2 v}$. When $d > \frac{1+r}{2} v$, $\frac{\partial \pi_1^D}{\partial r}$, $\frac{\partial \pi_2^D}{\partial r}$, and $\frac{\partial \Pi^D}{\partial r}$ are negative. $\frac{\partial \pi_1^D}{\partial d} = \frac{2d+v+rv}{4(1+r)v} > 0$ ; $\frac{\partial \pi_2^D}{\partial d} = \frac{2d+v+rv}{8(1+r)v} > 0$; $\frac{\partial \Pi^D}{\partial d} = \frac{3(2d+v+rv)}{8(1+r)v} > 0$. ∎

According to proposition 1, it is counterintuitive that all players' profits are negatively related with $r$ with the fact that the demand is positively related with $r$. The reason is that as $d$ increases, the negative impact of double marginalization (which can be shown as $\Pi^* - \Pi^D = \frac{(2d+v+rv)^2}{32(1+r)v}$) on the profits increases. As a result, when $d$ is sufficiently high the profits will actually decrease as $r$ increases.

As in the case of decentralized channel, we can get the equilibrium price and profit as shown in the Lemma 2.

***Lemma 2.*** *In the centralized data supply chain, the equilibrium price is $p^* = \frac{1}{4}(v + rv - 2d)$, the channel profit is $\Pi^* = \frac{(v+rv+2d)^2}{8(1+r)v}$; and the consumer surplus is $CS^* = \frac{(2d+v+rv)^2}{16(1+r)v}$.*

Since the proof is similar to that of Lemma 1, we skip it.

## 3.2 Nash Bargaining Pricing Model

Instead of directly charging the application provider data usage fee ($w$), the data provider can share $\alpha$ fraction of the application price without charging any upfront fee; this pricing mechanism is widely adopted in information goods (e.g., App) and platform business model (Chen, Fan & Li, 2016). As it is the case with Iyer and Villas-Boas (2003), the channel members make the price decisions through bargaining. In order to obtain analytical solutions, we adopt Nash Bargaining solution (Nash 1950, Binmore, Rubinstein & Wolinsky, 1986, Ghosh & Shah, 2015) and we call this pricing mechanism Nash Bargaining pricing (denoted by superscript $B$). Under this pricing mechanism, the expected profit functions of the data provider, the application provider, and the data supply chain are as follows:

$$\pi_1^B = (\alpha p + d)\frac{(1+r)v - 2p}{(1+r)v} \tag{3}$$

$$\pi_2^B = (1-\alpha)p\frac{(1+r)v - 2p}{(1+r)v} \tag{4}$$

$$\Pi^B = \pi_1^B + \pi_2^B = (p + d)\frac{(1+r)v - 2p}{(1+r)v} \tag{5}$$

We investigate two bargaining models. In the first one (denoted by superscript $B1$), we consider a case in which the upstream data provider and downstream application provider bargain on the sharing ratio $\alpha$, and then, given the sharing ratio, the application provider decides the application price. In the second model (denoted by superscript $B2$), the data provider and application provider bargain on both sharing ratio $\alpha$ and application price $p$.

### 3.2.1 Bargaining on Sharing Ratio

In this bargaining model, we use the backward induction to solve the two-stage game. In the second stage, the application provider, given the sharing ratio $\alpha$, decides application price to maximize its profit.

$$\max_p \pi_2^{B1} = (1-\alpha)p\frac{(1+r)v - 2p}{(1+r)v} \tag{6}$$

$$\text{s.t. } 0 < \frac{(1+r)v - 2p}{(1+r)v} \leq 1.$$

The application provider's optimal pricing is $p^{B1} = \frac{1}{4}(1+r)v$, which satisfies the demand constraint ($\frac{(1+r)v-2p}{(1+r)v} = \frac{1}{2}$).

In the first stage, the data provider and the application provider decide the sharing ratio by adopting Nash Bargaining model, which is given in the following objective function:

$$\max_{\alpha} \ \pi_1^{B1} \pi_2^{B1} = (\alpha p + d)\frac{(1+r)v-2p}{(1+r)v}(1-\alpha)p\frac{(1+r)v-2p}{(1+r)v} \tag{7}$$

$$\text{s.t. } 0 < \alpha < 1$$

Solving for the first-order condition, we get $\alpha = \frac{v+rv-4d}{2(1+r)v}$. It can be shown that the second-order condition is satisfied ($\frac{\partial^2 \ \pi_1^{B1}\pi_2^{B1}}{\partial \alpha^2} = -\frac{(1+r)^2 v^2}{32} < 0$). When $v > \frac{4d}{1+r}$, $0 < \alpha < 1$. As a result, we can obtain the following lemma.

**Lemma 3.** *Under the condition $v > \frac{4d}{1+r}$, when bargaining on sharing ratio only, the optimal price and sharing ration are $p^{B1} = \frac{1}{4}(1+r)v$, $\alpha^{B1} = \frac{v+rv-4d}{2(1+r)v}$; the expected profits of data provider, application provider and data supply chain are $\pi_1^{B1} = \frac{1}{16}(4d + v + rv)$, $\pi_2^{B1} = \frac{1}{16}(4d + v + rv)$, $\Pi^{B1} = \frac{1}{8}(4d + v + rv)$; the consumer surplus is $CS^{B1} = \frac{1}{16}(1+r)v$.*

Based on Lemma 3, we get the following proposition.

***Proposition 2.*** *The data provider has no incentive to hide the information of feedback data value in the bargaining, even though it can get a higher sharing ratio of the total profit in so doing.*

**Proof.** If the data provider hides the information of feedback data value (in that case $d = 0$), we can get its sharing ratio $\alpha = \frac{1}{2} > \alpha^{B1} = \frac{v+rv-4d}{2(1+r)v}$, but the expected profit of data provider is $\pi_1 = \frac{1}{16}(1+r)v < \pi_1^{B1} = \frac{1}{16}(4d + v + rv)$. ∎

### 3.2.2 Bargaining on both sharing ratio and price

The data provider and application provider bargaining on both sharing ratio $\alpha$ and price $p$ is described by the following maximization problem.

$$\max_{p,\alpha} \pi_1^{B2}\pi_2^{B2} = (\alpha p + d)\frac{(1+r)v-2p}{(1+r)v}(1-\alpha)p\frac{(1+r)v-2p}{(1+r)v} \tag{8}$$

$$\text{s.t.} \quad 0 < \alpha < 1$$

$$0 < \frac{(1+r)v-2p}{(1+r)v} \leq 1$$

Solving this optimization problem, we get the following lemma.

**Lemma 4.** Under the condition $v > \frac{6d}{1+r}$, the unique equilibrium is $\alpha^{B2} = \frac{6d-v-rv}{4d-2v-2rv}$, $p^{B2} = \frac{v+rv-2d}{4}$; the expected profits of data provider, application provider and data supply chain are $\pi_1^{B2} = \frac{(2d+v+rv)^2}{16(1+r)v}$, $\pi_2^{B2} = \frac{(2d+v+rv)^2}{16(1+r)v}$, $\Pi^{B2} = \frac{(2d+v+rv)^2}{8(1+r)v}$; the consumer surplus is $CS^{B2} = \frac{(2d+v+rv)^2}{16(1+r)v}$.

**Proof.** By solving for first-order condition, we get three stationary points ($\alpha = \frac{6d-v-rv}{4d-2v-2rv}, p = \frac{v+rv-2d}{4}$), ($\alpha = 1, p = -d$) and ($\alpha = 1, p = 0$). Obviously, the second and third points are not maximum points. Then we prove ($\alpha^{B2} = \frac{6d-v-rv}{4d-2v-2rv}$, $p^{B2} = \frac{v+rv-2d}{4}$) is the maximum point. With $v > \frac{6d}{1+r}$, $\left.\frac{\partial^2 \pi_1^B \pi_2^B}{\partial \alpha^2}\right|_{(\alpha^{B2},p^{B2})} = -\frac{(v+rv-2d)^2(2d+v+rv)^2}{32(1+r)^2v^2} < 0$;

$\left.\frac{\partial^2 \pi_1^B \pi_2^B}{\partial p^2}\right|_{(\alpha^{B2},p^{B2})} = -\frac{(2d+v+rv)^2(12d^2-4d(1+r)v+(1+r)^2v^2)}{4(1+r)^2v^2(v+rv-2d)^2} < 0$; $\left.\frac{\partial^2 \pi_1^B \pi_2^B}{\partial \alpha \partial p}\right|_{(\alpha^{B2},p^{B2})} = \frac{d(2d+v+rv)^2}{4(1+r)^2v^2} > 0$. Then we get $\left.\frac{\partial^2 \pi_1^B \pi_2^B}{\partial \alpha^2} \cdot \frac{\partial^2 \pi_1^B \pi_2^B}{\partial p^2} - (\frac{\partial^2 \pi_1^B \pi_2^B}{\partial \alpha \partial p})^2\right|_{(\alpha^{B2},p^{B2})} = \frac{(v+rv-2d)^2(2d+v+rv)^4}{128(1+r)^4v^4} > 0$. Therefore, ($\alpha^{B2} = \frac{6d-v-rv}{4d-2v-2rv}$, $p^{B2} = \frac{v+rv-2d}{4}$) is the maximum point of $\pi_1^{B2}\pi_2^{B2}$. And $v > \frac{6d}{1+r}$ ensures $0 < \alpha^{B2} < 1$ and $0 < \frac{(1+r)v-2p^{B2}}{(1+r)v} < 1$. Substituting $\alpha^{B2} = \frac{6d-v-rv}{4d-2v-2rv}$, $p^{B2} = \frac{v+rv-2d}{4}$ we get the expected profits and consumer surplus. ∎

Based on Lemma 4, we compare the two bargaining structures and get the following proposition.

*Proposition 3 (a) Bargaining on both sharing ratio and price dominates the bargaining only on sharing ratio, and it can achieve the supply chain's optimal performance; (b)*

the upstream data provider and downstream application provider share the supply chain profit at fixed ratio $\frac{1}{2}$.

Proposition 3 shows that bargaining only on sharing ratio can distort the price decision, leaving the supply chain performance suboptimal. However, if the marginal profit is equal for upstream and downstream ($d = 0$ in our setting), the two bargaining structures have the same performance ($\pi_1^{B1} = \pi_2^{B1} = \pi_1^{B2} = \pi_2^{B2} = \frac{v(1+r)}{16}$), indicating that the feedback data from end-users makes the two bargaining structures different.

## 3.3 Comparison of the Pricing Mechanisms

In this section, we compare the decentralized pricing with Nash Bargaining pricing from the perspectives of data provider, application provider, and end-users respectively.

***Proposition 4.*** *Data supply chain members' preferences for pricing mechanisms are as follows:*

a) *For data provider, decentralized pricing is equivalent to bargaining on both sharing ratio and price, and they both dominate bargaining only on sharing ratio.*

b) *For application provider, end-users, and data supply chain as a whole, bargaining on both sharing ratio and price dominates bargaining only on sharing ratio, which dominates decentralized pricing.*

**Proof**. For data provider, $\pi_1^D - \pi_1^{B1} = \frac{d^2}{4v+4rv} > 0$, $\pi_1^D - \pi_1^{B2} = 0$, and $\pi_1^{B1} - \pi_1^{B2} = -\frac{d^2}{4v+4rv} < 0$. For application provider, $\pi_2^D - \pi_2^{B1} = -\frac{4d(1+r)v+(1+r)^2v^2-4d^2}{32(1+r)v} < 0$, $\pi_2^D - \pi_2^{B2} = -\frac{(2d+v+rv)^2}{32(1+r)v} < 0$, and $\pi_2^{B1} - \pi_2^{B2} = -\frac{d^2}{4v+4rv} < 0$. For end-users, $CS^D - CS^{B1} = \frac{4d^2+4d(1+r)v-3(1+r)^2v^2}{64(1+r)v} < 0$ when $v > \frac{2d}{1+r}$, $CS^D - CS^{B2} = -\frac{3(2d+v+rv)^2}{64(1+r)v} < 0$, and $CS^{B1} - CS^{B2} = -\frac{d(d+v+rv)}{4(1+r)v} < 0$. For the data supply chain as a whole, $\Pi^D - \Pi^{B1} = -\frac{4d(1+r)v+(1+r)^2v^2-12d^2}{32(1+r)v} < 0$ when $v > \frac{2d}{1+r}$, $\Pi^D - \Pi^{B2} = -\frac{(2d+v+rv)^2}{32(1+r)v} < 0$, and $\Pi^{B1} - \Pi^{B2} = -\frac{d^2}{2v+2rv} < 0$. ∎

Proposition 4 shows the conflict of the preferences for pricing mechanisms between players in the data supply chain. Specifically, since data provider is the price leader in the decentralized pricing, it prefers decentralized pricing. This is anecdotally supported by the fact that the data providers usually poste data price on the website (http://www.datatang.com). However, other players in the data supply chain are worse off under decentralized pricing. Furthermore, although bargaining on both sharing ratio and price can achieve the optimal performance in the data supply chain, it allocates the maximum profit equally between upstream and downstream players. As a result, it lacks the flexibility of profit allocation. In the next section, we will address how to coordinate data pricing perfectly.

## 4. Coordinating Pricing with Revenue Sharing

Revenue sharing contract (Cachon & Lariviere, 2005) differs from the Nash Bargaining pricing in that it needs the application provider to pay the upfront data usage fee $w'$. Under revenue sharing contract (denoted by superscript $R$), the expected profit functions of data provider, application provider, and data supply chain are as follows.

$$\pi_1^R = (\rho p + w' + d)\frac{(1+r)v - 2p}{(1+r)v} \tag{9}$$

$$\pi_2^R = ((1-\rho)p - w')\frac{(1+r)v - 2p}{(1+r)v} \tag{10}$$

$$\Pi^R = \pi_1^R + \pi_2^R = (p + d)\frac{(1+r)v - 2p}{(1+r)v} \tag{11}$$

***Proposition 5.*** *When the revenue sharing contract parameter satisfies $w' = (\rho - 1)d$ ($0 < \rho < 1$), the data supply chain are perfectly coordinated, that is, the data provider and application provider can share the maximum supply chain profit at any ratio.*

**Proof.** Substituting $w' = (\rho - 1)d$ to the profit functions, we can directly get $\pi_1^R = \rho(p+d)\frac{(1+r)v-2p}{(1+r)v} = \rho\ \Pi^R$, and $\pi_2^R = (1-\rho)(p+d)\frac{(1+r)v-2p}{(1+r)v} = (1-\rho)\ \Pi^R$. The profit functions of data provider and application provider are affine transformations of the data supply chain's profit function. Therefore, individual player's optimal pricing also maximizes the expected profit of the data supply chain. ∎

Proposition 5 suggests that, different from the traditional revenue sharing contract, the wholesale price $w' = (\rho - 1)d$ is negative because $0 < \rho < 1$. As a result, the data provider needs to subsidize the application provider. The reason is that the data provider can get the feedback data from end-users. Thus, The revenue sharing mechanism provides a win-win pricing strategy for all the player in the data supply chain.

## 5. Concluding Remarks

In the emerging big data industry, data product is different from both traditional product and information goods. By focusing on new features of data products, this paper investigates data pricing mechanism from the data supply chain perspective. This paper studies three pricing mechanisms' performance and their effects on the participants in the data industry. Using game theoretical model, we obtain analytical solutions in each pricing mechanism which includes the decentralized and centralized pricing, Nash Bargaining pricing, and revenue sharing mechanism. First, we find that decentralized pricing has the lowest performance. Second, although Nash Bargaining pricing can achieve the centralized channel performance, the upstream data provider and downstream application provider can only equally divide the total channel profit. Third, revenue sharing, in which the data provider subsidizes the application provider, can achieve the first best performance and divide the maximum profit arbitrarily. Accordingly, end-users benefit mostly from the bargaining pricing and revenue sharing. The findings provide further understanding of the pricing mechanism in data transaction and offer some guidelines for the mechanism design of data pricing in the big data industry.

There are some limitations in this research. First, we only consider a Nash Bargaining solution in a symmetric setting for tractability issue. Future research can extend the bargaining to an asymmetric setting. Second, in order to focus on the pricing mechanism and to capture the profit allocation problem in this mechanism, we only consider the price decision in the data supply chain and simplify the dynamic process of the data usage. Further research can consider a process of complex data usage.